\documentclass[preprint,aps,floatfix,superscriptaddress,prb,showpacs]{revtex4}
\usepackage{amsmath, amsthm, amssymb, graphicx}
\input epsf.sty
\begin{document}

\def\g{\gamma}
\def\r{\rho}
\def\w{\omega}
\def\wo{\w_0}
\def\wp{\w_+}
\def\wm{\w_-}
\def\t{\tau}
\def\av#1{\langle#1\rangle}
\def\pf{P_{\rm F}}
\def\pr{P_{\rm R}}
\def\F#1{{\cal F}\left[#1\right]}

\title{Equilibrium and non-equilibrium properties of synthetic metamagnetic films: A Monte Carlo study}

\author{James Mayberry, Keith Tauscher and Michel Pleimling}
\affiliation{Department of Physics, Virginia Tech, Blacksburg, Virginia 24061-0435, USA}

\date{\today}

\begin{abstract}
Synthetic antiferromagnets with strong perpendicular anisotropy can be modeled by layered Ising antiferromagnets.
Accounting for the fact that in the experimental systems the ferromagnetic layers, coupled antiferromagnetically
via spacers, are multilayers, we propose a description through Ising films where ferromagnetic stacks composed
of multiple layers are coupled antiferromagnetically. We study the equilibrium and non-equilibrium properties of these
systems where we vary the number of layers in each stack. Using numerical simulations, we 
construct equilibrium temperature$-$magnetic field phase diagrams for a variety of cases. We find the same dominant features
(three stable phases, where one phase boundary ends in a critical end point, whereas the other phase boundary shows
a tricritical point at which the transition changes from first to second order) for all studied cases.
Using time-dependent quantities, we also study the ordering processes that take place after a temperature quench.
The nature of long-lived metastable states are discussed for thin films, whereas for thick films we compute
the surface autocorrelation exponent.

\end{abstract}


\maketitle

\section{Introduction}

Heterostructures formed by ferromagnetic multilayers that are coupled antiferromagnetically via spacers
have a wide range of possible applications, ranging from high-density storage technology to spintronic
devices.\cite{Full03,Mang06,Mukh09,Egg11} Cases with strong perpendicular anisotropy (as encountered, 
for example, in [Co/Pt]/Ru or [Co/Pt]/NiO with ferromagnetic [Co/Pt] multilayers) have been
the objects of intensive studies, both experimentally \cite{Hell03,Hell07,Hau08,Liu09,Kise10a,Che13} and 
theoretically.\cite{Ross04a,Ross04b,Ross04c,Kise08,Kise10a,Kise10b,Kise11} 
Due to the overall structure of these antiferromagnetically
coupled multilayers, they are sometimes referred
to as synthetic or artificial metamagnets. 
In phenomenological approaches, these systems are commonly modeled by layered antiferromagnets where a ferromagnetic
multilayer is described by a single variable, namely the total magnetization of the
multilayer. For strong perpendicular anisotropy, this naturally
leads to the description via an Ising metamagnet where layers with ferromagnetic in-layer interactions 
are coupled antiferromagnetically. Ising bulk metamagnets have been the subject of many
studies in the past and their properties are well understood by now.\cite{Harb73,Kinc75,Land81,Herr82,Land86,
Herr93,Selk95,Selk96,Pleim97,Gala98,Sant98,Zuko00,Sant00,More02} 
Surprises show up, however, when studying Ising metamagnets in thin film geometry. In systems
with an even number of layers and a magnetic field applied perpendicular to the surfaces,
the temperature-field phase diagram exhibits a new phase
in addition to the two phases of the bulk system (the low field antiferromagnetic and the high field paramagnetic phases):
for intermediate field strengths, the surface layer, which is magnetized oppositely to the applied magnetic
field in the ground state, aligns with the magnetic field. 
The phase transition between the antiferromagnetic phase and this intermediate 
phase is discontinuous and ends in a critical end point.\cite{Chou11}

The characterization of the ferromagnetic multilayers by only their overall magnetization
is reasonable at low temperatures where thermal
fluctuations are irrelevant. However, at higher temperatures, thermal properties will change with the width
of the ferromagnetic stacks.
In this paper, we address how the extent of these stacks changes the
properties of Ising metamagnetic films.
Motivated by the desire to better model the 
experimental synthetic metamagnets with ferromagnetic multilayers, 
we numerically study the thermal properties of 
films composed of ferromagnetic stacks that are coupled antiferromagnetically. Using large-scale numerical
simulations, we determine the phase diagrams for systems with stacks of different sizes. A comparison of different 
cases reveals many common features as well as quantitative differences.

In addition to this study of equilibrium properties, we also investigate relaxation processes.
Non-equilibrium processes in systems undergoing phase ordering, including domain growth and aging phenomena, have
been the source of continuous interest in the last decades. 
The majority of studies were either restricted to simple
ferromagnets,\cite{Bray94,Cugl03,Cala05,Henk10} yielding a rather comprehensive understanding of these processes in this type
of systems, or to more complex situations where glassy dynamics result from competition and
frustration effects,\cite{Kaw04,Henk07} such as in the case of spin glasses. 
However, similar processes also take place in other systems and the
metamagnets discussed in this work provide an interesting example. A further motivation
for studying relaxation processes in the layered antiferromagnets comes from an experimental study
that observed aging phenomena in Co/Cr superstructures where the Cr spacer layers
provide an antiferromagnetic exchange coupling between the ferromagnetic Co films.\cite{Mukh10}
Since Co films have an in-layer anisotropy, the Co/Cr superstructures are not well described by the models
discussed here. Still, the observation of magnetic metastability and of slow dynamic relaxation in these superstructures
is very intriguing and reveals a need to better understand ordering and relaxation processes in layered antiferromagnets.
In the second part of this paper, we present first steps in the study of non-equilibrium processes encountered in 
our systems during a temperature quench. After preparing the system in a disordered state, 
corresponding to high temperature, we bring it in contact
with a heat bath at a temperature at which the equilibrium system is ordered. This temperature quench may be
done in the presence or absence of an applied external magnetic field. Besides analyzing thin films,
we also consider the non-equilibrium surface properties of thick films in order to extract
the value of the surface autocorrelation exponent.

The paper is organized in the following way: In the next Section we introduce the models under investigation.
Section III is devoted to a comprehensive study of the equilibrium properties of metamagnetic Ising films 
where ferromagnetic stacks are coupled antiferromagnetically. We discuss the changes in the temperature-magnetic field phase
diagram that result from varying the thickness of the ferromagnetic stacks. In Section IV we study some aspects of non-equilibrium
relaxation taking place after a temperature quench. 
We conclude in Section V.

\section{Models}

\begin{figure} [h]
\includegraphics[width=0.65\columnwidth]{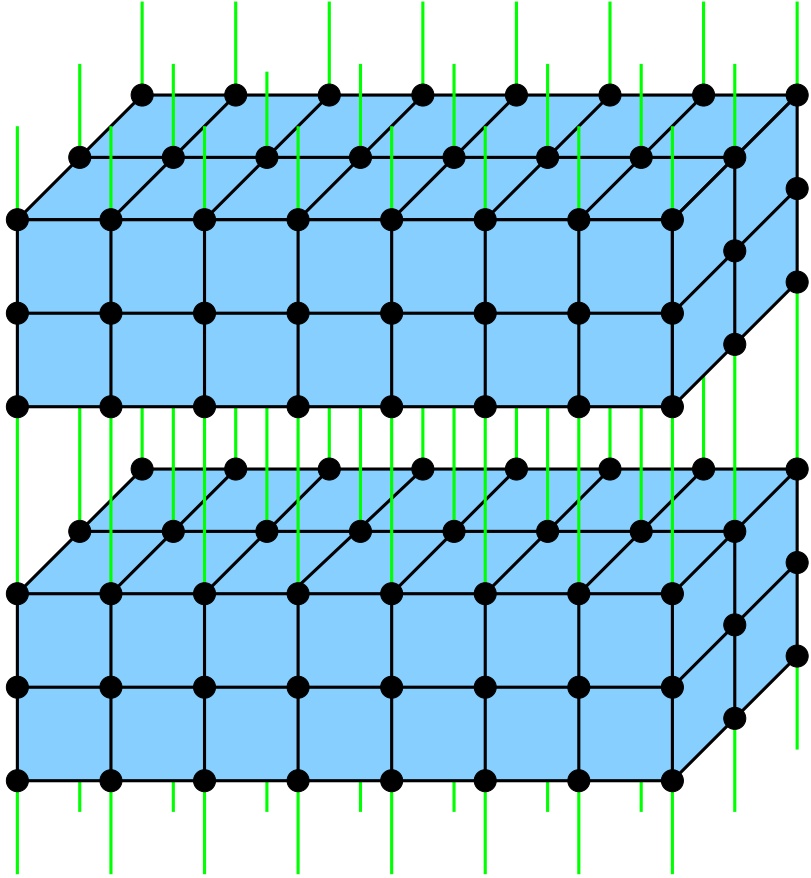}
\caption{\label{fig1} (Color online)
Sketch of a system where ferromagnetic stacks formed by $n=3$ layers are coupled antiferromagnetically.
The black and green bonds indicate ferromagnetic and antiferromagnetic couplings respectively.}
\end{figure}

In the following, we consider Ising films where every lattice site $(x,y,z)$ is characterized by an Ising spin
$s(x,y,z) = \pm 1$. We realize the film geometry by considering $N$ layers along the $z$-direction, using free
boundary conditions in that direction so that the layers at $z=1$ and $z=N$ form surfaces. 
In the other two directions we have periodic boundary conditions. Our layered systems
are composed of ferromagnetic stacks that are coupled antiferromagnetically, see Fig. \ref{fig1} for an example. 
In the case of single ferromagnetic layers, this
yields the standard Ising metamagnet. In our study, we allow for stacks of varying sizes $n$. The
results discussed in the following have been obtained for systems with $n=1$, 2, 3, and 4.

The number of layers $n$ in a stack and the number of stacks $N/n$ in the films studied in 
this work are close to
those found in experimental systems. For example for [Co/Pt]/Ru the number of Co/Pt repeats per stack is 
$1 < n < 14$, whereas the number of stacks in a film is in the range $1 < N/n < 20$.\cite{Hell07}

We can write the Hamiltonian of the system as
\begin{equation} \label{eq:ham}
{\mathcal H} = - J \sum\limits_{in~layer} s(x,y,z) s(x',y',z) - \sum\limits_{z=1}^{N-1} J_z
\sum\limits_{x,y} s(x,y,z) s(x,y,z+1) - h \sum\limits_{x,y,z} s(x,y,z)~,
\end{equation}
where the third term, which is the product of the magnetic field with strength $h >0$ and the magnetization
$M = \sum\limits_{x,y,z} s(x,y,z)$, follows when applying a magnetic field pointing in the positive-$z$ direction.
The first term is the ferromagnetic in-layer nearest neighbor coupling with strength $J > 0$, 
whereas the second term describes the
interaction between spins located in two consecutive layers. These two terms together
form the internal energy $E$. The coupling constant $J_z$ can be either
positive, yielding a ferromagnetic coupling between layers in the same stack, or negative, yielding
an antiferromagnetic coupling between stacks. For the example shown in Fig. \ref{fig1},
we have $J_z < 0$ when $z$ is a multiple of 3 and $J_z > 0$ otherwise. For simplicity, we consider
in this work only the case $\left| J_z \right| = J$ where $J$ is the strength of the in-layer interaction.

The ground state phase diagrams as a function of magnetic field strength are readily obtained. 
We first note that we need to distinguish between cases where the number of ferromagnetic stacks is odd
or even. For an odd number of stacks (each composed of $n$ layers), we have only two phases: 
the low magnetic field phase where the
stacks are coupled antiferromagnetically and the high field paramagnetic phase where all spins are
aligned with the field. The transition between these phases takes place when the
field strength $h = 2 J/n$. As an example, consider the case with $n=2$ and $N=6$. If $h < J$, then
the stable phase is $++--++$ where the signs represent the signs of the magnetization in the
different layers. For $h > J$ the stable phase is given by $++++++$. The more interesting situation
is that of a system composed by an even number of stacks. In that case, the phase diagram displays
an intermediate phase where both surfaces and the stacks they belong to align with the field. For example, 
for the case $n=2$ and $N=8$ the phase sequence for increasing magnetic field is $++--++--$ 
$\longrightarrow$ $++--++++$ $\longrightarrow$ $++++++++$. The transitions take place at $h = J/n$
and $h = 2 J/n$. In the following, we focus on situations with an even number of stacks.

\section{Equilibrium properties}

In order to clarify the thermal, equilibrium properties of the different systems we compute
a variety of global quantities through extensive Monte Carlo simulations. 
These quantities include the average magnetization density
\begin{equation}
m = \langle M \rangle/V~,
\end{equation}
the average energy density
\begin{equation}
e = \langle E \rangle/V~,
\end{equation}
the magnetic susceptibility
\begin{equation}
\chi = \frac{1}{V T} \left( \langle M^2\rangle  - \langle M \rangle^2 \right)~,
\end{equation}
and the specific heat
\begin{equation}
c  = \frac{1}{V T^2} \left( \langle E^2\rangle - \langle E \rangle^2 \right)~. 
\end{equation}
Here $V$ is the total number of sites in the system, whereas $T$ is the temperature measured
in units such that $J/k_B=1$. The brackets in these equations indicate both a time and an ensemble
average. The time average is done after the system has reached equilibrium.

\begin{figure} [h]
\includegraphics[width=0.80\columnwidth]{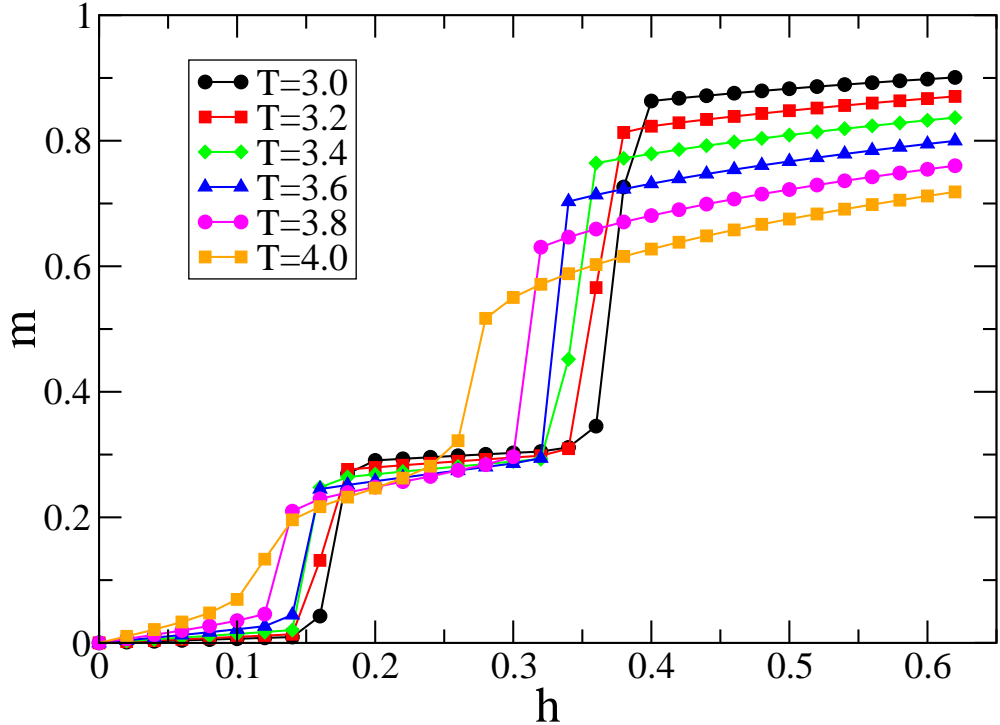}
\caption{\label{fig2} (Color online) The magnetization density as function of the magnetic field
strength for a system with stacks containing four ferromagnetically coupled layers. The system
is composed of $80 \times 80 \times 24$ spins. At low magnetic fields a first partial ordering takes place,
followed by the alignment of all the layers with the field for a larger value of $h$. 
Data for different temperatures are shown.
Here and in the following error bars are smaller than the symbol sizes.
}
\end{figure}

We simulate the systems using the heat-bath algorithm.\cite{Landau} As usual the time unit is one Monte Carlo
step (MCS) during which, on average, every spin is considered once for an update.

In order to relax rapidly to equilibrium we prepare the system in an initial state
which has the same magnetization profile as the $h=0$ ground state but with a small 
magnetization density (for our 
production runs we used a density of $\pm 0.04$). The advantages of this initial state are two-fold.
First, as discussed in the next section, when starting from a fully disordered state
the relaxation time to reach equilibrium can be very large, as the system can get stuck in
long-lived metastable states. Second, when starting from one of the ground states with fully magnetized
layers, it becomes very difficult to flip whole layers deep inside the ordered phase. 
As a result it is not easy 
to obtain reliable data close to the phase transitions.
We found that our initial state is a good compromise that yields in each case a fast relaxation towards the
equilibrium state.

\begin{figure} [h]
\includegraphics[width=0.80\columnwidth]{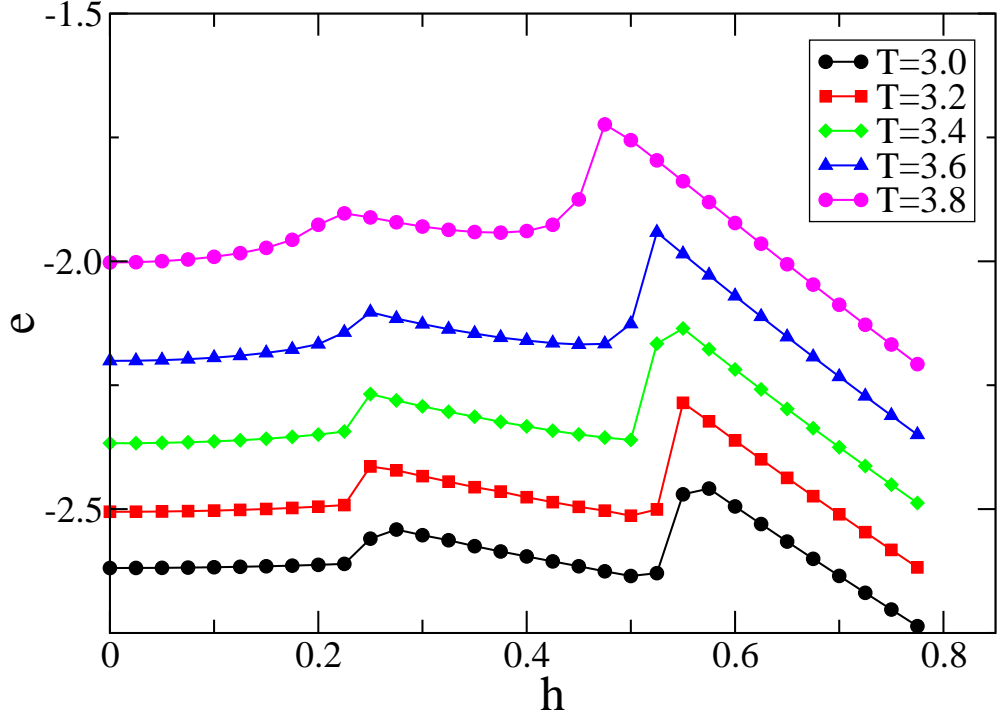}
\caption{\label{fig3} (Color online) The energy density as function of the magnetic field strength for
a three layers per stack system. The size of the system is $80 \times 80 \times 18$. The two successive transitions
show up as sudden increases followed by a decrease of the energy with increasing field strength.
}
\end{figure}

We consider films of different thicknesses that contain an even number of stacks of width $n$ 
(with $n=1$, 2, 3, or 4).
The layers within the stacks are composed of $K \times K$ spins, with $K$ ranging from 20 to 320.
After having reached equilibrium we take a time average over at least 50000 MCS. The data discussed in the 
following result from averaging over an ensemble of typically 100 independent runs.

\begin{figure} [h]
\includegraphics[width=0.80\columnwidth]{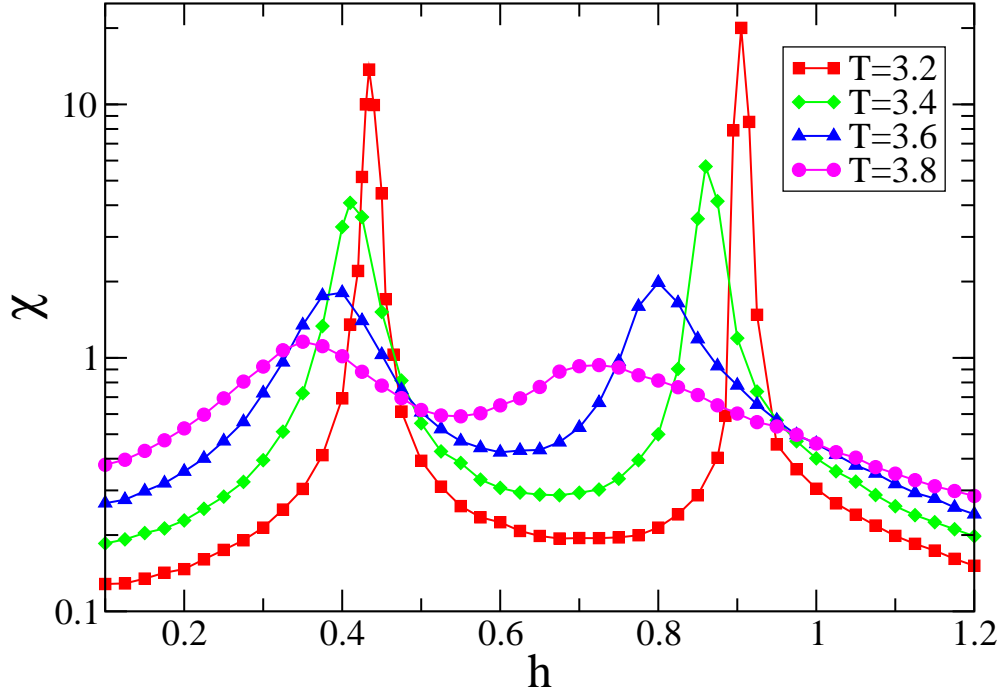}
\caption{\label{fig4} (Color online) The magnetic susceptibility versus the magnetic field strength for a
system where every stack contains two layers. The system is composed of $20 \times 20 \times 8$ spins.
}
\end{figure}

Figs. \ref{fig2}-\ref{fig4} show examples of the data obtained in our simulations. Fig.\ \ref{fig2}
displays the magnetization density as a function of the magnetic field strength in a system of $N= 24$
layers with $n=4$ layers in each ferromagnetically coupled stack. The two phase transitions, from the
antiferromagnetic arrangement of the stacks with vanishing total magnetization to the intermediate phase
and from the intermediate phase to the paramagnetic phase where all layers are aligned with the magnetic field,
are clearly visible through the step-like changes in the magnetization. The values of $m$ in the intermediate
phase and in the paramagnetic phase depend on the temperature. Whereas the steps are very abrupt
for small temperatures, at higher temperatures they become much smoother, due to an increase of thermal
fluctuations. As shown in Fig.\ \ref{fig3}
for a system with $N=18$ and $n=3$, both transitions provide a similar clear signal when plotting the
energy density as a function of $h$. These changes also show up as peaks in the susceptibility $\chi(T,h)$ 
(see Fig.\ \ref{fig4} for an example with $N=8$ and $n=2$) and in the specific heat $c(T,h)$.

In order to obtain the temperature$-$magnetic field phase diagrams for fixed values of $N$ and $n$,
we analyze data for systems with layer sizes ranging from $20 \times 20$ to $320 \times 320$.
Finite-size scaling, where we compare systems with different layer sizes,
allows us to take into account the shift of the positions of the maxima
in the response functions (the susceptibility and the specific heat) with the lateral size of the system.

\begin{figure} [h]
\includegraphics[width=0.80\columnwidth]{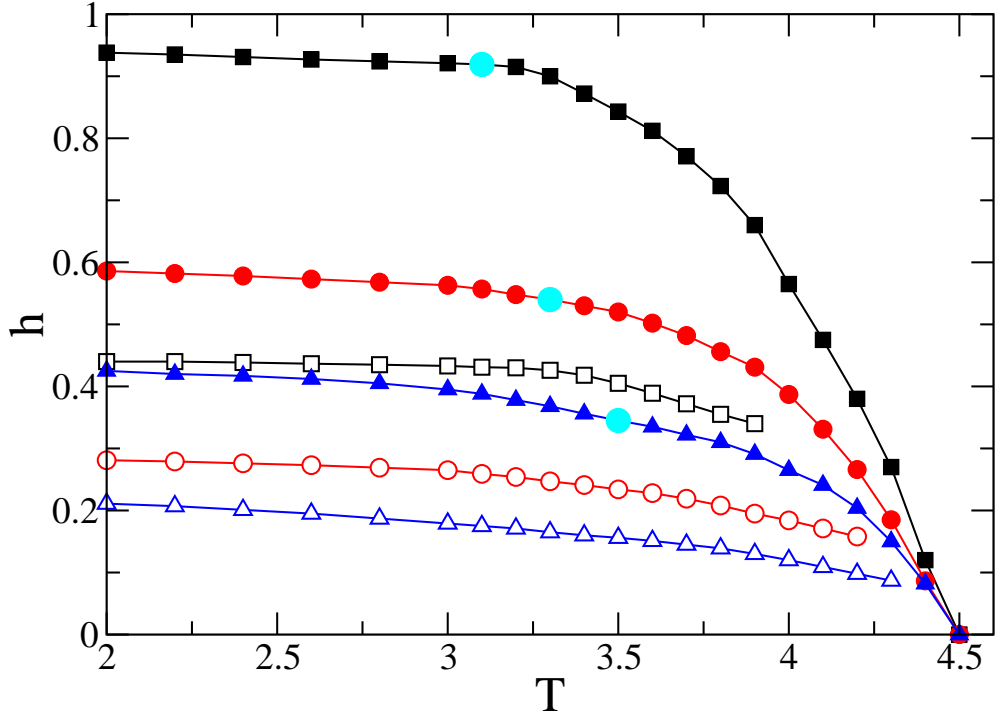}
\caption{\label{fig5} (Color online) Temperature$-$magnetic field phase diagrams for three different systems:
$N=16$ and $n=2$ (black squares), $N=18$ and $n=3$ (red circles), $N=24$ and $n=4$ (blue triangles). The open
symbols indicate the phase transition between the antiferromagnetically aligned stacks at low fields and the
intermediate phase with a partial alignment to the magnetic field. For all studied cases this transition is
discontinuous and ends in a critical end point. The lines with closed symbols show the phase transitions to the
high field paramagnetic phase. The large cyan (grey) circles indicate the tricritical points
separating the discontinuous transition at low temperatures from the continuous transition at high temperatures.
The phase diagrams have been obtained through a finite-size scaling analysis of the susceptibility
and specific heat for systems with layer
sizes ranging from $20 \times 20$ to $320 \times 320$ spins.
}
\end{figure}

Fig. \ref{fig5} summarizes our results of the finite-size scaling analysis of the susceptibility
and specific heat for three cases, namely a system with 2-layer stacks that form a film
of width $N=16$ (black squares), a system that contains six stacks containing three layers (red circles), and
a system with six stacks of four layers (blue triangles). The phase diagrams 
for all three cases have the same qualitative features
(which are similar\cite{Chou11} to those encountered in the phase diagram of the metamagnetic film with $n=1$),
with three phases (the phase with an antiferromagnetic ordering of the stacks
at low fields, the intermediate phase with alignment
of one additional surface surface with the magnetic field for intermediate field strengths, and the paramagnetic phase for
large fields) filling the phase diagram. The fields at which the transitions take place strongly depend
on the number $n$ of layers in the ferromagnetic stacks, in agreement with the exactly known values
at $T=0$, see the discussion in Sect. II. 
At low temperatures all transitions are discontinuous, whereas
at higher temperatures the transition to the paramagnetic phase changes its character and becomes continuous.
The cyan (grey) circles in Fig. \ref{fig5} indicate our estimations for the location of the tricritical
points where this change takes place. The transition lines at smaller fields, which separate the phase
with antiferromagnetically coupled stacks from the intermediate phase, keep their discontinuous character
until they end in a critical end point. Increasing the size of the stacks extends this transition line to higher
temperatures and lower fields. Already for our system with four layers in each stack, the lower transition line 
approaches closely the paramagnetic transition line before it stops (see open blue triangles in Fig. \ref{fig5}).

\section{Non-equilibrium relaxation}
In order to monitor the non-equilibrium processes, we prepare the system in a disordered
initial state (corresponding to infinite temperatures). Then, the system is quenched deep into the ordered phase
by bringing it in contact with a heat bath at the corresponding temperature. This protocol assures that 
no correlations are present at the beginning of the ordering process. In an experimental situation, 
some correlations will build up even during a very rapid quench, but this is a complication that we
do not consider in the following.

Having prepared our system in this way, we let it evolve with time. A good understanding of the non-equilibrium
processes can only be achieved when looking at local, i.e. layer-dependent, quantities. For that reason
we focus in the following on the time-dependent magnetization in each layer, given by
\begin{equation}
M(z,t) = \sum\limits_{x,y} s(x,y,z;t)~.
\end{equation}
In addition to analyzing individual runs, we also extract the typical behavior from 
the average time-dependent planar magnetization density
\begin{equation}
m(z,t) = \overline{M(z,t)}/A~,
\end{equation}
where $\overline{\cdots}$ indicates the ensemble average and $A = K \times K$ is the number of sites in each layer.

We also study the spin-spin autocorrelation in each layer, defined by
\begin{equation}
C(z,t) = \overline{\sum\limits_{x,y} s(x,y,z;t) s(x,y,z;0)}/A
\end{equation}
which compares configurations at time $t$ with the initial state. The surface autocorrelation can then be
obtained by averaging over the surfaces located at $z=1$ and $z=N$:
\begin{equation} \label{surfcorr}
C_s(t) = \left[ C(1,t) + C(N,t) \right]/2~,
\end{equation}
whereas for the thick films the corresponding bulk quantity can be extracted from the middle
of the sample:
\begin{equation} \label{bulkcorr}
C_b(t) = \left[ C(N/2,t) + C(N/2+1,t) \right]/2~.
\end{equation}

\subsection{Ordering processes}

\begin{figure} [h]
\includegraphics[width=0.90\columnwidth]{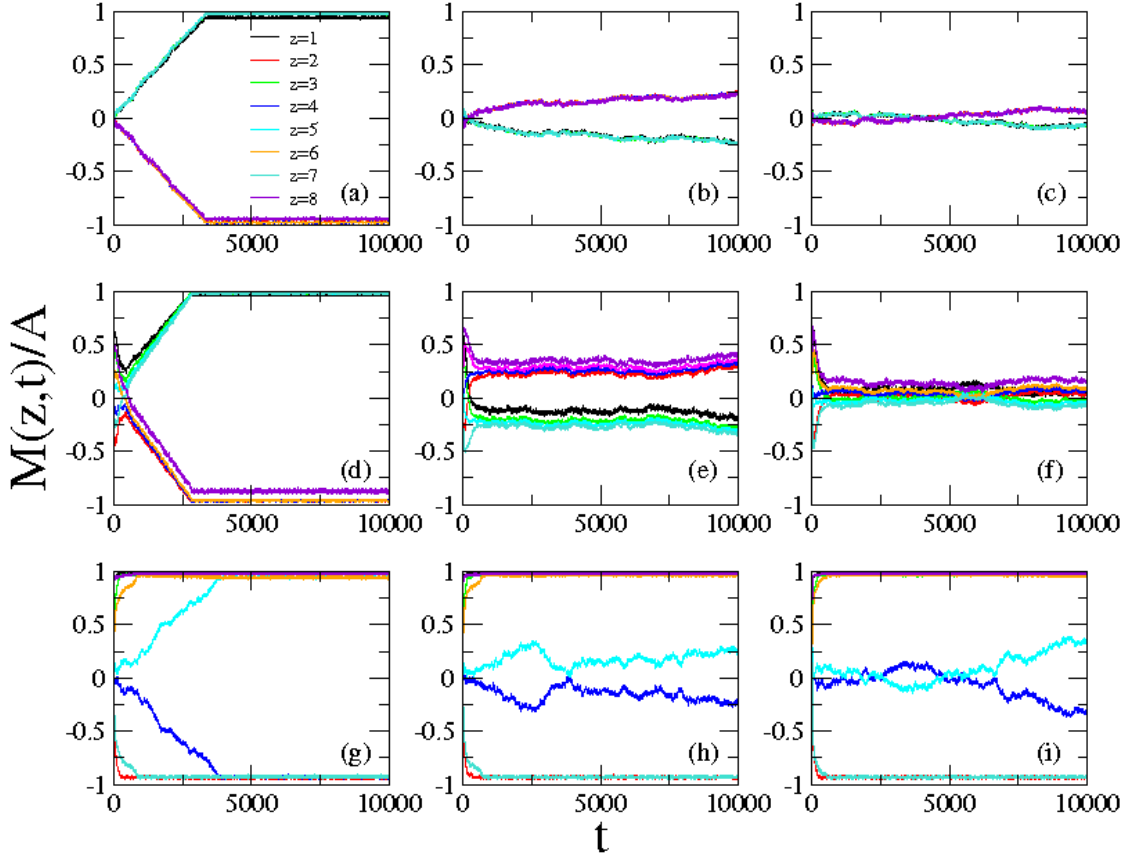}
\caption{\label{fig6} (Color online)
Temporal evolution of the layer magnetization densities for a metamagnetic film
($n=1$) composed of 8 layers. Whereas the left column presents runs where the system rapidly relaxes
toward equilibrium, the middle and right columns show cases where the system gets trapped in a long-lived
metastable state. The magnetic field strength is (a)-(c) $h=0$, (d)-(f) $h=0.6$,
and (g)-(i) $h=1.1$, with the temperature being held constant at $T=2.5$. The system is prepared
in a disordered initial state. Each layer contains $80 \times 80$
spins.
}
\end{figure}

\begin{figure} [h]
\includegraphics[width=0.90\columnwidth]{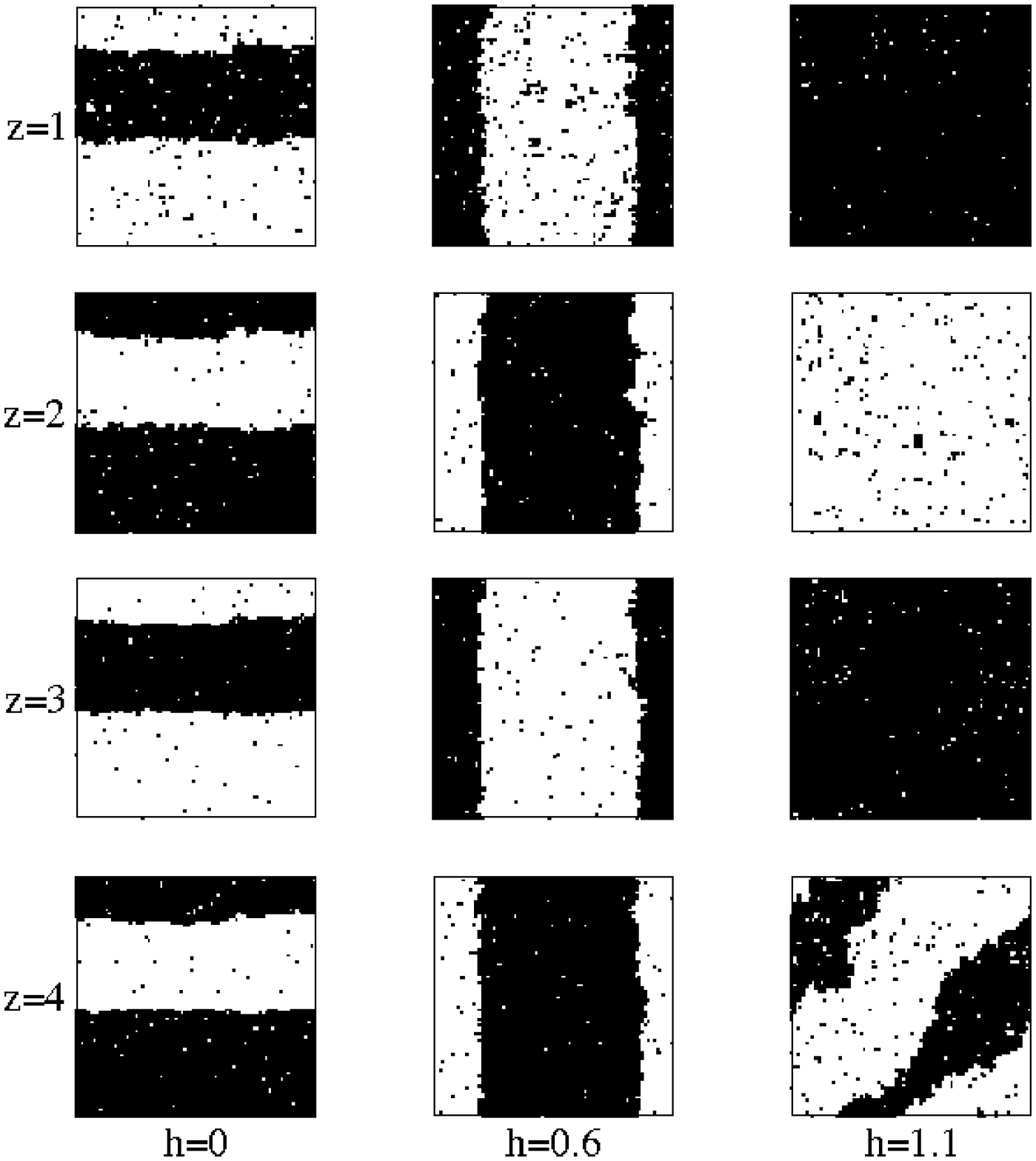}
\caption{\label{fig7} 
Spin configurations in the layers 1 to 4 at the end of some of the runs shown in Fig.\ \ref{fig6}
for which the system has not yet reached equilibrium after t=10000 MCS.
Left column: panel (b) in Fig.\ \ref{fig6}, middle column: panel (e), right column: panel (h).
Spins with value $+1$ are shown as black squares.
}
\end{figure}

Preparing a system in an initially disordered (high temperature)
state before quenching it down to below the critical temperature
is a protocol which is often used, both experimentally \cite{Mukh10} and theoretically \cite{Henk10}, 
in the study of ordering processes and aging phenomena
in magnets. In simple ferromagnets, small domains are formed immediately after the quench. It is generally thought
that at a later
stage, during the coarsening process, the larger domains grow at the expense of the smaller ones until the 
system ends up in one of the stable equilibrium states.
However, a series of papers on zero temperature coarsening in the Ising model
revealed that this is a too naive picture,\cite{Spir01,Spir02,Barr09,Olej11a,Olej11b} 
as the system often ends up in an infinitely long-lived
metastable state (in three space dimensions this is in fact the generic situation \cite{Olej11a,Olej11b}). 
At higher temperatures the
system will eventually find its way out of these metastable states and end up in equilibrium,
but the time spent in these states can be very large. 

In the following, we discuss the ordering processes in metamagnetic films exposed to an external magnetic field.
The competition between the antiferromagnetic inter-stack coupling and the magnetic field adds another
level of complexity to the coarsening process.
We focus on the cases with $n=1$ and $n=2$ as no fundamentally new aspects emerge for the larger values of $n$.

Figs.\ \ref{fig6} and \ref{fig7} give examples of our results for the metamagnetic films with antiferromagnetic coupling
between neighboring layers ($n=1$). The data in these figures have been obtained for films with $N=8$ layers
and a fixed temperature of $T=2.5$. 
Besides the case without applied magnetic field, we also consider field strengths of
$h=0.6$, still inside the low field antiferromagnetic phase, and $h=1.1$, where the system is in the intermediate phase.

Panels (a), (d), and (g) in Fig.\ \ref{fig6} show the time evolution of the layer magnetization for the generic case where
the system rapidly (within a few thousand MCS) relaxes and ends in an equilibrium state.
However, in roughly 35\% of the cases the system is caught in a metastable state from which it escapes only
after a very long time. For every value of the external field we provide examples of runs with long-lived
metastable states in the right two panels of Fig. \ref{fig6}. 

For $h=0$ this observation of long-lived metastable states is in agreement with the known properties of the Ising model.
Indeed, inspection of the Hamiltonian (\ref{eq:ham}) in absence of a magnetic
field, i.e. with $h=0$, reveals that changing
the signs of the antiferromagnetic couplings and the signs of every second layer at the same time
yields the Hamiltonian of the standard three-dimensional Ising model. Because of that, the properties of the metamagnets
in a vanishing magnetic field are identical to those of the corresponding Ising model. This symmetry is of course 
broken when adding a magnetic field. For $h=0.6$ the time evolution of the planar magnetizations is similar to
that at $h=0$. As $h > 0$ in this case, 
there is a bias towards positive magnetizations as minus spins flip more easily than plus spins.
However, the long-lived configurations are of the same type as those found for $h=0$, see the spin configurations shown in
the first two columns of Fig.\ \ref{fig7}. As long as the antiferromagnetic phase is the equilibrium phase, the long-lived
metastable phase results from the presence of
very straight and stable domain walls that separate two domains formed by the two equilibrium states. 

\begin{figure} [h]
\includegraphics[width=0.90\columnwidth]{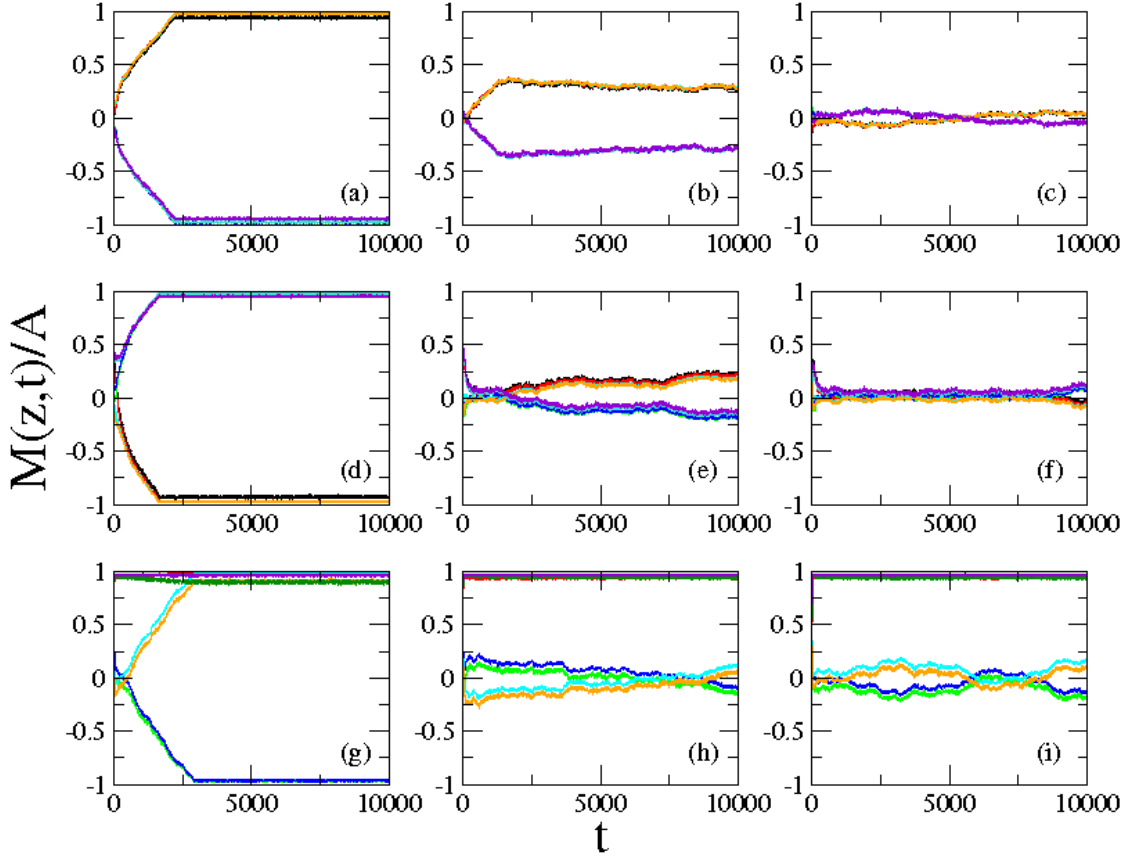}
\caption{\label{fig8} (Color online)
Temporal evolution of the layer magnetization densities for a film of $N=8$ layers composed of four
two-layer stacks ($n=2$). Whereas the left column presents runs where the system rapidly relaxes
toward equilibrium, the middle and right columns show cases where the system gets trapped in a long-lived
metastable state.
The magnetic field strength is (a)-(c) $h=0$, (d)-(f) $h=0.2$,
and (g)-(i) $h=0.55$, with the temperature being held constant at $T=2.5$. The system is prepared
in a disordered initial state. Each layer contains $80 \times 80$
spins.
}
\end{figure}

For $h=1.1$ and $T=2.5$, see the last row in Fig. \ref{fig6} and the last column in Fig. \ref{fig7}, the equilibrium phase 
is the intermediate phase where both surface layers are aligned with the magnetic field,
resulting in five layers with positive and three layers with negative magnetization.\cite{Chou11}
The metastable phases, see panels (h) and (i) in Fig. \ref{fig6}, are those where in the middle two layers, with
$z=4$ and $z=5$, domain walls separate positively and negatively magnetized domains. See the $z=4$ configuration
in the right column of Fig. \ref{fig7}. It is interesting to note that these domain walls are not straight
but instead are oriented predominantly in the diagonal direction. 
The resulting configuration, $+-+\cdot\cdot+-+$, can be explained by the observation
(shown in the time evolution in panels (g)-(i) in Fig. \ref{fig6})
that the ordering starts at the two surfaces and progresses from there into the film.
Starting from the outside, the system might end up in either the state $+-+-++-+$ or the state $+-++-+-+$, so that the
two middle layers can either have positive or negative magnetization.
As a result, it can happen that in these layers long-lived domain walls form due to the competition
between the two equilibrium phases,
yielding the observed metastable states.

Finally, for large values of $h$ where the paramagnetic phase prevails, the system rapidly forms positively magnetized
layers, and no long-lived metastable phases are observed.

\begin{figure} [h]
\includegraphics[width=0.90\columnwidth]{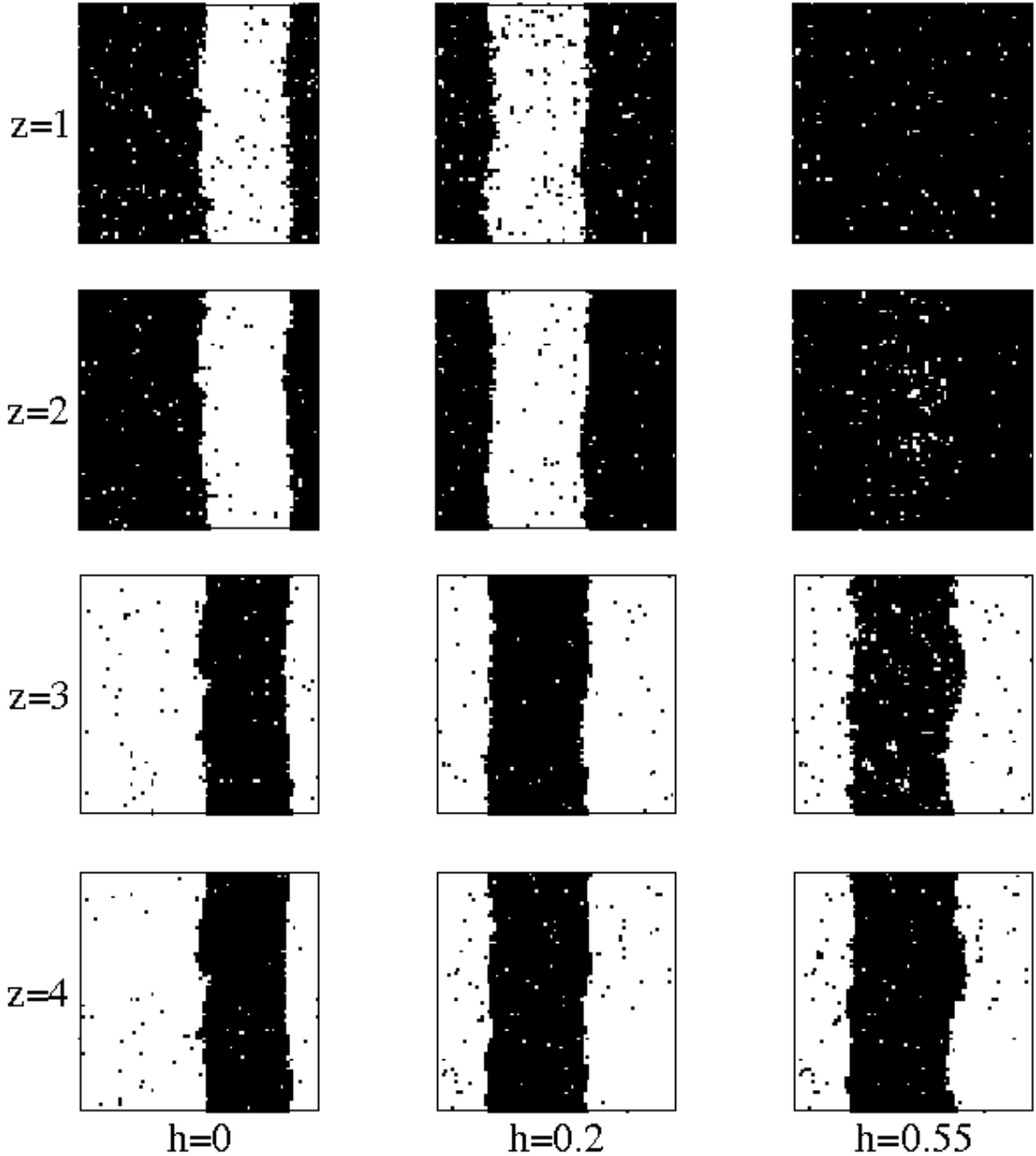}
\caption{\label{fig9} 
Spin configurations in the layers 1 to 4 at the end of some of the runs shown in Fig.\ \ref{fig8}
for which the system has not yet reached equilibrium after t=10000 MCS.
Left column: panel (b) in Fig.\ \ref{fig8}, middle column: panel (e), right column: panel (h).
Spins with value $+1$ are shown as black squares.
}
\end{figure}

Figs. \ref{fig8} and \ref{fig9} show the results of a similar study for systems with $N=8$ layers
where each ferromagnetic stack contains two layers ($n=2$). Overall the phenomenology is very similar
to that of the $n=1$ systems. In the low field phase (as encountered for $h=0$ and $h=0.2$) 
the metastable states are again characterized by straight domain walls. The only apparent difference
comes from the two-layer structure of the ferromagnetic stacks
and can be seen by comparing Fig. \ref{fig7} and Fig. \ref{fig9}.
As seen in Fig. \ref{fig8} (g)-(i) and the last 
column in Fig. \ref{fig9}, inside the intermediate phase, the mechanism yielding metastable states
is the same as for the $n=1$ case. The ordering grows from the outside to the inside,
yielding a competition between the states $++--++++$ and $++++--++$. As a result, long-lived domain walls might form in
the four innermost layers, resulting in the observed metastable states. Contrary to the $n=1$ case where the domain walls
are pointed mainly in the diagonal direction, see the third column in Fig. \ref{fig7}, the $n=2$ case
(as well as other cases with $n > 1$) presents domain walls that are predominantly oriented in axial direction.

\subsection{Surface autocorrelation exponent}
Whereas in the previous discussion we focused on thin films, we now consider the situation of very thick
films such that the film thickness is much larger than the bulk correlation length. In that case, the two surfaces
are uncorrelated and the system can be considered to model the behavior of a semi-infinite system. This allows
us to extract properties of the surface layer during the relaxation process without having to worry about
possible effects due to the presence of the second surface.

Relaxation processes close to surfaces have been studied in a few instances in the past. Most of these studies
focused on quenches to the critical point of a magnetic system.\cite{Rits95,Maju96,Cala05,Plei04,Baum06,Marc12}
In \cite{Baum07} the two-dimensional semi-infinite Ising model quenched below the critical point was studied 
and the local relaxation processes close to the surface were investigated with the help of correlation and response
functions. 

Systems with coarsening are characterized by a growing correlation length (typical domain length) $L(t)$. In the
regime where $L(t)$ is large compared to the microscopic length scales but small compared to the system size, the bulk
autocorrelation (\ref{bulkcorr}) decays algebraically with $L(t)$:
\begin{equation} \label{bulkcorr2}
C_b(t) \sim \left(L(t) \right)^{- \lambda_b}~,
\end{equation}
where $\lambda_b$ is the bulk autocorrelation exponent.\cite{Fish88,Huse89,Henk10}
For the two-dimensional Ising model, it was found that a surface locally changes this behavior, yielding a 
surface autocorrelation (\ref{surfcorr}) that decays in the same regime as
\begin{equation} \label{surfcorr2}
C_s(t) \sim \left(L(t) \right)^{- \lambda_s}~,
\end{equation}
with a surface autocorrelation exponent $\lambda_s > \lambda_b$.\cite{Baum07}
We measured the autocorrelation in our systems in order to see whether a similar relationship holds
in three dimensions. The symmetry of the Hamiltonian (\ref{eq:ham}) in absence of a magnetic
field yields the Hamiltonian of the standard three-dimensional Ising model when changing 
the signs of the antiferromagnetic couplings and the signs of the spins in every other stack of 
ferromagnetically coupled layers at the same time. This symmetry makes sure
that the autocorrelation exponents measured in the synthetic metamagnets with $h=0$ are the same as those of the
three-dimensional Ising model. We verified this by also directly simulating the three-dimensional Ising
model, see below.

In our simulations, we consider large systems consisting of $100 \times 100 \times 80$ spins. The dimensions
of our samples are large enough to avoid the appearance of finite-size effects during the non-equilibrium
relaxation process. Data of high 
quality are needed in order to be able to reliably extract values for the autocorrelation exponents from
time-dependent correlation functions. The data discussed in the following result from averaging over at least 75000
independent runs.

We carefully checked that during the simulation the time-dependent correlation length $L(t)$, extracted from 
the exponentially decaying
spatial correlations, remained much smaller than the extent of the system. We also verified that 
the correlation length in our systems increases as the square root of time, $L(t) \sim t^{1/2}$, as expected for a system
with curvature driven coarsening.\cite{Brow02}

\begin{figure} [h]
\includegraphics[width=0.80\columnwidth]{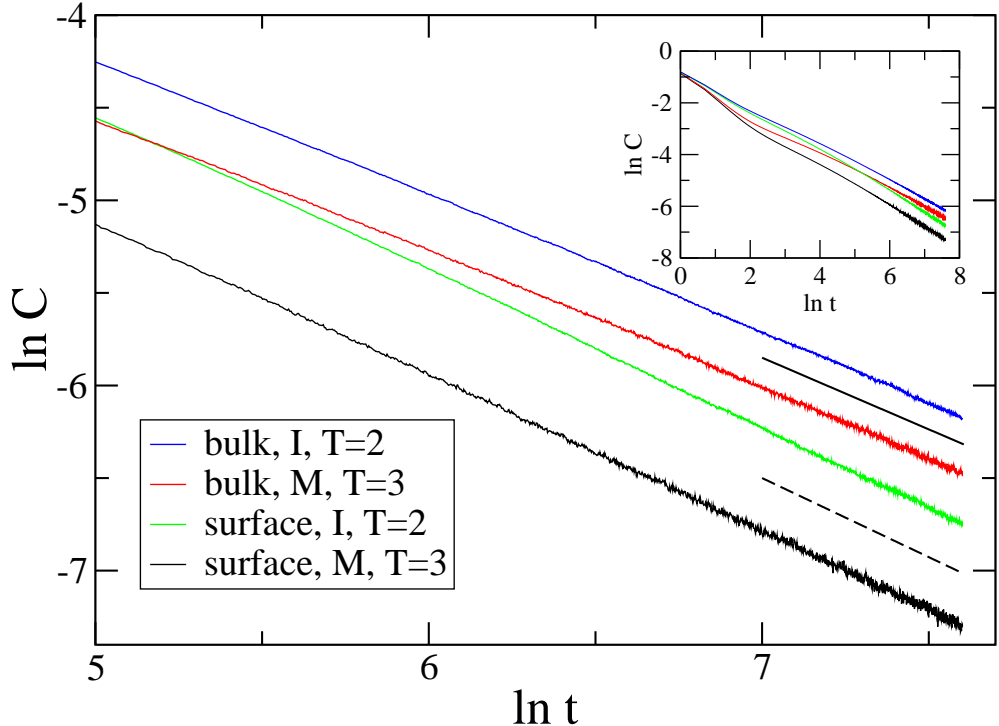}
\caption{\label{fig10} (Color online) 
Time-dependent bulk and surface autocorrelation in absence of a magnetic field
for the metamagnet ($n=1$) and the three-dimensional
Ising model quenched inside the ordered phase. 
The slopes in the log-log plot yield the ratio between the bulk resp. surface autocorrelation exponent and the
dynamic exponent $z$ (i.e. $\lambda_b/z$ resp. $\lambda_s/z$).
The solid line indicates the slope 0.774, whereas the dashed line has the slope 0.854. 
The size of the simulated systems is $100 \times 100 \times 80$.
}
\end{figure}

Fig. \ref{fig10} shows the temporal evolution of the autocorrelation at the surface and in the bulk for the layered
antiferromagnet with $n=1$ at $T=3$ as well as for the three-dimensional Ising model at $T=2$. Due to the
difference in temperature, differences are observed in the early time behavior, see the inset of Fig. \ref{fig10}. 
However, in the long time limit,
the slopes for the different quantities are found to be the same for the two situations under investigation.
Thus for the bulk autocorrelation exponent we find for the metamagnet $\lambda_b = 1.550(6)$, whereas for the Ising
model we have $\lambda_b = 1.546(6)$. Our estimates for $\lambda_b$ refine the values of 1.59 resp. 1.60(2)
obtained in earlier studies
of non-equilibrium relaxation in the three-dimensional Ising model.\cite{Huse89,Henk03}
For the surface autocorrelation exponent, for which no previous estimates are available, we obtain
the values $\lambda_s = 1.706(8)$ for the metamagnet and $\lambda_s = 1.712(8)$ for the Ising model. We note that
$\lambda_s > \lambda_b$, so that the situation is similar to that encountered in the two-dimensional Ising 
model.\cite{Baum07} However, the values for the surface and bulk exponents are much closer in three than in
two dimensions.

\section{Conclusion}
Bulk metamagnets have been the topic of intensive research for many years, and, by now, their properties are  
well understood. The recent emergence of artificial antiferromagnets in the form of superstructures
in thin film geometry has resulted in a need to also better understand the properties of metamagnetic thin films.
In a realistic description one should take into account the fact that artificial antiferromagnets are composed of ferromagnetic
multilayers, with an antiferromagnetic exchange coupling between the multilayers provided by the 
spacer layers. 

In this paper, we took into account the finite extent of these multilayers and used an Ising model where ferromagnetic
stacks composed by a certain number of layers are coupled through an antiferromagnetic inter-stack interaction.
Using large scale numerical simulations, we studied stacks of different sizes and investigated both their equilibrium
and non-equilibrium properties. 

The equilibrium phase diagrams as a function of temperature and applied magnetic field have the same qualitative
feature for any size of the stacks, albeit quantitative differences are observed. 
As already found for the standard Ising metamagnet in thin film geometry,\cite{Chou11}
a third phase, called the intermediate phase, is stabilized due to the finiteness of the film. In this phase, both surfaces
(and the stacks they belong to) align with the magnetic field, whereas inside the film an antiferromagnetic ordering of the
stacks still prevails. The transition between the low field antiferromagnetic phase and the intermediate phase
is discontinuous, and the transition line ends in a critical end point. The second phase transition, from the
intermediate phase to the paramagnetic phase stable at high fields, is of first order at low temperatures and
of second order at high temperatures. The existence of a tricritical point is a feature that is common to the
films and the bulk systems.

The non-equilibrium relaxation is complicated by the competition between antiferromagnetic
coupling and the magnetic field. Whereas in most cases the system relaxes rapidly to the equilibrium state, in
roughly 35\% of the cases with low and intermediate field strengths the system ends up in a long-lived 
metastable state. The nature of the metastable state can be understood through the analysis of the time-dependent
planar magnetization as well as through the inspection of typical spin configurations. Starting from a disordered
initial state, the surfaces order first. As the ordering progresses from the outside to the inside,
competition between different spin configurations can take place in the middle of the sample yielding
long-lived domain walls in the innermost layers. We note that for $n=1$ these domain walls are predominantly in
the diagonal direction. This is an unexpected result, and a more in-depth investigation, especially at very low
temperatures, will be needed to gain a better understanding of the mechanism behind this.

Finally, we studied thick films in absence of a magnetic field
in order to compute the surface autocorrelation function. We found
that this quantity decays algebraically, with the surface autocorrelation exponent taking on a value
that differs from that of its bulk counterpart. Due to the symmetry of the Hamiltonian for vanishing magnetic
fields, this exponent is the same as that for the three-dimensional Ising model.

\begin{acknowledgments}
This work is supported by the US National
Science Foundation through grant DMR-1205309. We thank Dr. Yen-Liang Chou for his early contributions
to this project.
\end{acknowledgments}

\end{document}